\documentclass[fleqn,10pt,twocolumn]{TexTemplate/wlscirep}
\usepackage[utf8]{inputenc}
\usepackage[T1]{fontenc}
\usepackage{booktabs}
\usepackage{makecell}
\usepackage[table]{xcolor}
\usepackage{caption}
\definecolor{headergray}{gray}{0.92}
\definecolor{rowgray}{gray}{0.97}

\title{Human adaptive variability stabilises collective traffic dynamics}

\author{
Shirui Zhou$^{1,2,5,\#}$,
Ching Jin$^{3,\#}$,
Junfang Tian$^{1,2,*}$,
Shiteng Zheng$^{4}$,
Rui Jiang$^{4,*}$,
Shiquan Zhong$^{1,2}$,
Shoufeng Ma$^{1,2}$,
Vittorio Loreto$^{5,6,7,*}$
\\
$^{1}$Institute of Systems Engineering, College of Management and Economics, Tianjin University, No. 92 Weijin Road, Nankai District, 300072, Tianjin, China.\\
$^{2}$Laboratory of Computation and Analytics of Complex Management Systems (CACMS), Tianjin University, No. 92 Weijin Road, Nankai District, 300072, Tianjin, China.\\
$^{3}$Centre for Interdisciplinary Methodologies, University of Warwick, Coventry, CV4 7AL, United Kingdom.\\
$^{4}$School of Systems Science, Beijing Jiaotong University, No. 3 Shangyuancun, Haidian District, 100044, Beijing, China.\\
$^{5}$Physics Department, Sapienza University of Rome, Piazzale Aldo Moro 5, Rome, RM, 00185, Italy.\\
$^{6}$Sony Computer Science Laboratories Rome, Joint Initiative CREF-SONY, Centro Ricerche Enrico Fermi, Via Panisperna 89/A, Rome, 00184, Italy.\\
$^{7}$Complexity Science Hub Vienna, Metternichgasse 8, Vienna, 1030, Austria\\
$^{\#}$These authors contributed equally to this work.\\
$^{*}$Correspondence should be addressed to J.T. (email: jftian@tju.edu.cn), 
R.J. (email: jiangrui@bjtu.edu.cn), 
or V.L. (email: vittorio.loreto@uniroma1.it).
}

\begin{abstract}
The rapid deployment of automated systems is often guided by the paradigm that replacing human behavioural variability with precise, uniform, algorithmic control inherently optimises collective performance. In socio-technical ecosystems like automotive traffic, this design philosophy underpins current commercial longitudinal control systems, such as adaptive cruise control (ACC). Here, by conducting two large-scale human-driving experiments (2.95 million car-following observations) ---a 25-vehicle platoon experiment and a controlled 11-driver protocol ---cross-validated against the publicly available NGSIM dataset (0.77 million observations) and an open repository of 22 production ACC systems, supplemented by empirically calibrated commercial-ACC simulations, we find a notable counter-effect: rigid algorithmic uniformity is associated with systemic fragility. Commercial rule-based control amplifies minor microscopic perturbations into macroscopic stop-and-go waves; in matched model simulations, this amplification is associated with approximately 2.7- to 5.0-fold increases in fuel consumption and carbon emissions across scenarios. Conversely, human-driven platoons progressively dissipate these disturbances, preserving smooth collective flow. We identify the behavioural mechanism underlying this human advantage, showing that human car-following does not obey a fixed, proportional spacing rule. Instead, human drivers continuously reshape their time-headway distributions across different speed regimes, consistent with a non-monotonic transition from efficiency-oriented to risk-sensitive regulation. This speed-dependent structural variability provides a plausible nonlinear damping mechanism that may suppress the synchronisation of local errors before they cascade system-wide. These findings suggest that the conventional engineering assumption---that human variability is merely suboptimal noise to be automated away---warrants re-examination. Beyond traffic dynamics, our work points to a design consideration for large-scale interactive AI systems: achieving macroscopic robustness may benefit from embedding adaptive, human-inspired behavioural flexibility rather than relying solely on rigid uniformity.
\end{abstract}

\begin{document}

\flushbottom
\maketitle

% \section{Introduction}

Automation is often understood as a way to replace human behavioural variability with precision, consistency, and algorithmic control. In transportation, this vision is especially compelling: autonomous and partially automated vehicles are expected to sense their surroundings more accurately than humans, respond more quickly, and execute explicit control laws with high repeatability. Their longitudinal control logic, as widely implemented in systems such as commercial adaptive cruise control (ACC), is typically engineered around predefined, proportional rules designed to guarantee localised tracking accuracy and predictability~\cite{helbingTrafficRelatedSelfdriven2001}. From this conventional engineering perspective, eliminating behavioural variability across a fleet should suppress individual error, rendering collective traffic flow smoother, safer, and inherently more stable~\cite{Shladover2018CAVOverview,Swaroop1994SpacingHeadway,Darbha1999ICCandStability,Wilson2011LinearStability}. Yet, macroscopic instability is a classic emergent collective phenomenon. Even in the absence of physical bottlenecks, tiny microscopic fluctuations in car-following can spontaneously self-organise into global stop-and-go waves—a structural vulnerability well-documented in canonical ring-road experiments and the statistical physics of traffic flow~\cite{helbingTrafficRelatedSelfdriven2001,Sugiyama2008}.

This discrepancy suggests that the paradigm of automation may rest on an oversimplified assumption: that individual-level variability is merely detrimental noise. While uniform, rigid responses simplify control in isolated engineering tasks, in complex interactive systems, uniformity is not synonymous with robustness~\cite{Simon1962ArchitectureComplexity,Page2010DiversityComplexity,Anderson1972MoreIsDifferent,Zhang2021RandomHeterogeneity}. On the contrary, a growing literature spanning collective problem solving, socio-technical governance, and complex network dynamics demonstrates that local heterogeneity in behavioural rules, rhythms, and cognitive repertoires is essential for sustaining systemic resilience. Crucially, when identical machine agents respond identically to identical inputs, their local errors can become globally synchronised and amplified. What appears as suboptimal noise or inconsistency at the level of the individual unit may therefore serve a vital stabilising function at the collective level by acting as a barrier against destructive synchronisation~\cite{Hong2004DiverseProblemSolvers,Baggio2019CognitiveDiversityCommons,Zhang2021RandomHeterogeneity,Meng2024PersonalisedStrategyUpdates}. Traffic networks provide an ideal testing ground for this principle, as macroscopic stability emerges directly from repeated microscopic human-algorithmic interactions. Indeed, empirical field data have recently shattered the assumption that production ACC improves traffic flow by default; across multiple commercial systems, rule-based automation exhibits brittle, string-unstable dynamics that transmit and magnify perturbations~\cite{Gunter2021,Makridis2020TRR,Ye2023TRR,Makridis2021,Ciuffo2021,Stern2018DissipationWaves}, whereas variable human drivers naturally act to damp them~\cite{Shi2021CommercialAVHeadway,tangComparingCarFollowingBehaviorPatterns2025,papadoulisEvaluatingSafetyImpactConnected2019,Gunter2021}. This behavioural contrast questions the expectation that rigid automation will inherently optimise collective safety and efficiency~\cite{abdel-atyMatchedCasecontrolAnalysisAutonomous2024}.

\begin{figure*}[!ht]
\centering
\includegraphics[width=0.95\linewidth]{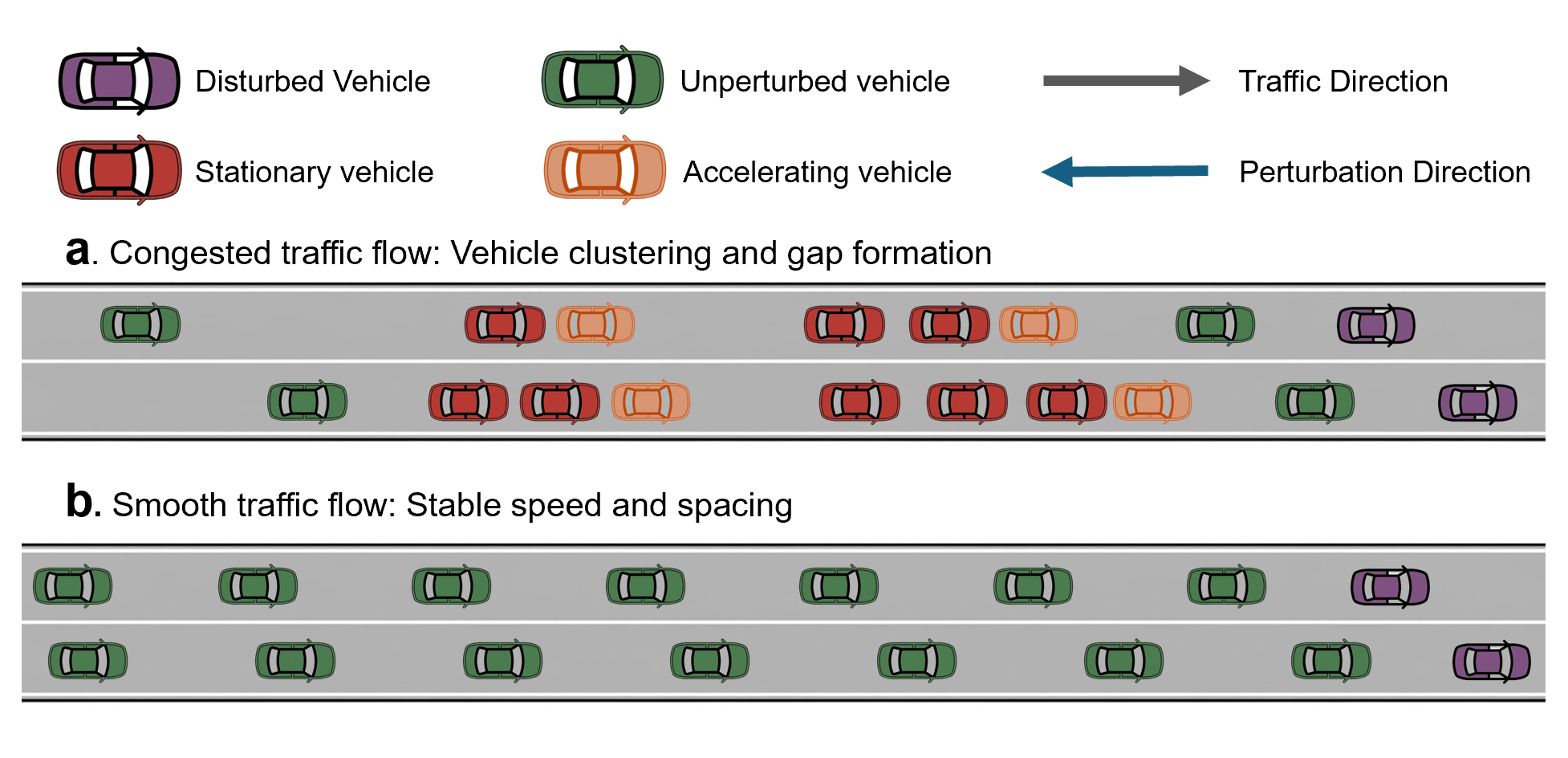}
\caption{
\textbf{Conceptual illustration of two contrasting modes of perturbation propagation in car-following.}
\textbf{(a)} In one mode, a small disturbance introduced by an upstream vehicle propagates backwards through the traffic stream, opposite to the direction of vehicle motion. As successive followers respond, the disturbance becomes increasingly organised: some vehicles slow sharply or stop, while others accelerate to close newly created gaps. This interaction produces alternating clusters and voids, corresponding to the emergence of stop-and-go structures.
\textbf{(b)} In the other mode, disturbances are progressively absorbed by successive followers through local adjustments in speed and spacing. As a result, fluctuations remain limited, no persistent clustering develops, and the traffic stream maintains comparatively smooth motion.
Vehicle colours denote disturbed, stationary, accelerating, and unperturbed states, and the arrows indicate traffic direction and perturbation direction. The schematic summarises the two qualitative regimes analysed in this study: perturbation amplification and perturbation damping.
}
\label{fig:concept_mechanism}
\end{figure*}

Figure~\ref{fig:concept_mechanism} schematically illustrates this behavioural contrast in perturbation dynamics. In the rigidly aligned regime, localised microscopic disturbances become synchronised and amplified as they propagate downstream, precipitating vehicle clustering, macro-gap formation, and structural stop-and-go waves. In the adaptively variable regime, these perturbations decay through heterogeneous local adjustments, allowing the collective traffic stream to maintain smooth, stable speed and spacing profiles. Under the conventional engineering expectation that automation optimises system performance through strict precision, uniformity, and control, commercially implemented autonomous driving should, in principle, favour the stabilising, adaptively variable regime. Whether this expectation holds true when rigid algorithms interact at scale, however, remains a critical empirical question.

This asymmetry raises a fundamental question about the behavioural foundations of stability in emergent socio-technical systems. Historically, human driving has been characterised as noisy, inconsistent, and suboptimal; empirical evidence demonstrates that drivers differ profoundly across individuals, and even a single operator rarely replicates their responses under identical external conditions~\cite{Jiang2014NatureCarFollowing,Jiang2015}. In traditional engineering paradigms, this behavioural variability is categorised as a systemic defect or residual noise that automation must systematically purge. However, in complex decentralised systems, behavioural variability can be deeply functional. Across biological, social, and technological domains, macro-stability often depends not on strict microscopic uniformity, but on a structured diversity of local responses. In cognitive groups, diversity expands the exploratory space for problem-solving strategies and reinforces the sustainable governance of common resources; similarly, in dynamic networks, heterogeneous local parameters preserve global coherence under severe feedback delays that completely destabilise homogeneous systems~\cite{Mason2012CollaborativeLearning,Rahwan2019}. In uncertain or dynamic environments, adaptive agents maximise utility by continually balancing exploitation of localised regularities with exploration of safety margins. Computational models of networked learning and collective information-seeking confirm that system-wide performance is optimised not by perfectly aligned, algorithmic repetition, but by response patterns that preserve local flexibility while still facilitating macro-coordination~\cite{Karpas2017InformationSocialtaxis}. What is conventionally dismissed as individual inconsistency may therefore constitute an evolved, adaptive mechanism that prevents catastrophic synchronisation at the collective level.

Here, however, our automated benchmark refers specifically to commercially implemented adaptive cruise control (ACC) and closely related rule-based longitudinal control paradigms, rather than the full cognitive stack of full vehicle autonomy. We therefore reserve broader terms such as automated driving for the broader technological context, using commercial ACC to describe the specific algorithmic architectures analysed here. By contrast, current commercial ACC systems are typically engineered around explicit and comparatively rigid longitudinal control laws. While these deterministic rules are highly effective at delivering precise target tracking and repeatable microscopic responses, they fundamentally lack the adaptive flexibility inherent to human behaviour. Consequently, despite possessing superior sensing metrics, higher computational bandwidth, and rapid actuation, these automated systems can remain remarkably brittle when complex collective conditions depart from the idealised steady-state assumptions embedded in their control design~\cite{koopmanAutonomousVehicleSafetyInterdisciplinary2017,rezaeiSafetyAutonomousVehiclesWhat2021,ibanez-guzmanLiDARCamerasAutonomousDriving2025,Ciuffo2021,hePhysicsaugmentedModelsSimulateCommercial2022}. The resulting divergence is not merely a superficial mismatch in reaction latencies or sensing thresholds; more fundamentally, it reflects a tension between two distinct modes of network coordination: adaptive regulation under systemic uncertainty and rule-based regulation under predefined, localised objectives. Resolving this paradigm distinction is essential for anticipating and mitigating the emergent, macro-scale consequences of automation.

In this study, we examine how these two divergent modes of control shape the macroscopic stability of interactive networks. Integrating controlled multi-vehicle car-following experiments, open-access production databases, and theoretical statistical analysis, we systematically isolate the collective consequences of human driving and commercially deployed automated control under closely matched conditions. Rather than treating individual human variability as an undesirable error or residual noise, we analyse it as a highly structured behavioural phenomenon, evaluating whether it provides a vital, stabilising counterweight to systemic shocks. Our results demonstrate that the performance gap between human and machine driving cannot be captured by superficial metrics such as localised sensing precision or raw actuation latency. Instead, it exposes a deeper systemic principle: individual adaptive variability functions as a mechanism for collective damping, whereas rigid algorithmic uniformity is associated with the synchronisation and growth of network disturbances. At the microscopic level, this divergence maps onto the structural contrast between context-sensitive human spacing strategies and the linear, constant-time-headway control logic that dominates current commercial architectures~\cite{cascettaAutonomousVehiclesDriveHumans2022,gunterModelBasedStringStabilityAdaptive2020}. More broadly, our findings suggest that the human collective advantage emerges not from stochastic randomness, but from a structured, velocity-dependent adaptation profile---an empirical basis for rethinking the engineering of resilient machine behaviour in large-scale socio-technical systems~\cite{Tian2019SpeedAdaptation,Li2021ACCBehavior,Swaroop1994SpacingHeadway,Dey2009DesiredTimeGap}.

More generally, our results speak to a profound, overarching question now emerging across the study of artificial systems: when large numbers of machine agents interact, macro-scale collective outcomes depend less on localised sensing accuracy or control precision, and far more on the emergent behavioural organisation induced by shared algorithmic rules. From this viewpoint, the traffic patterns unmasked here constitute a vivid, real-world manifestation of the broader machine-behaviour problem~\cite{Rahwan2019,Jennings2014HumanAgentCollectives}, illustrating how rigidly aligned, homogeneous local responses can unexpectedly generate large-scale, unintended collective effects. Beyond the domain of intelligent transportation, this finding suggests that robustness in complex distributed systems may depend on preserving controlled forms of behavioural diversity, which allow local units to remain sensitive and adaptive to shifting macroscopic conditions. Engineering autonomous systems under the assumption that eliminating individual variability optimises the network may miss an important source of systemic resilience. A far more effective and pressing paradigm shift is to incorporate adaptive, context-sensitive flexibility directly into automated control laws~\cite{Dafoe2021CooperativeAI,Jennings2014HumanAgentCollectives}, ensuring that deployment at scale enhances collective stability rather than systemically undermining it \cite{Rahwan2019}.
\begin{figure*}[htbp]
\centering
(a)\includegraphics[width=0.47\linewidth]{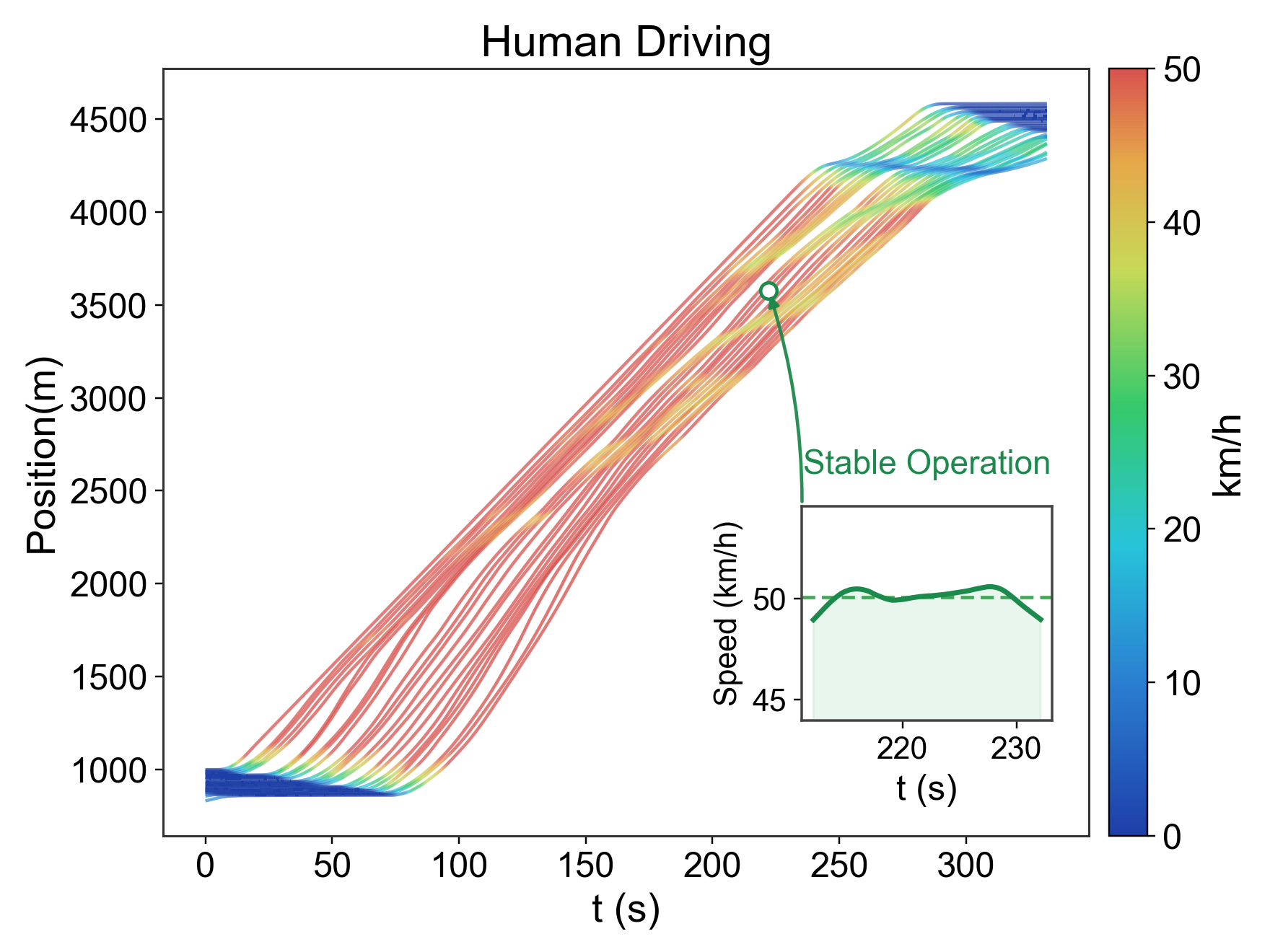}
(b)\includegraphics[width=0.47\linewidth]{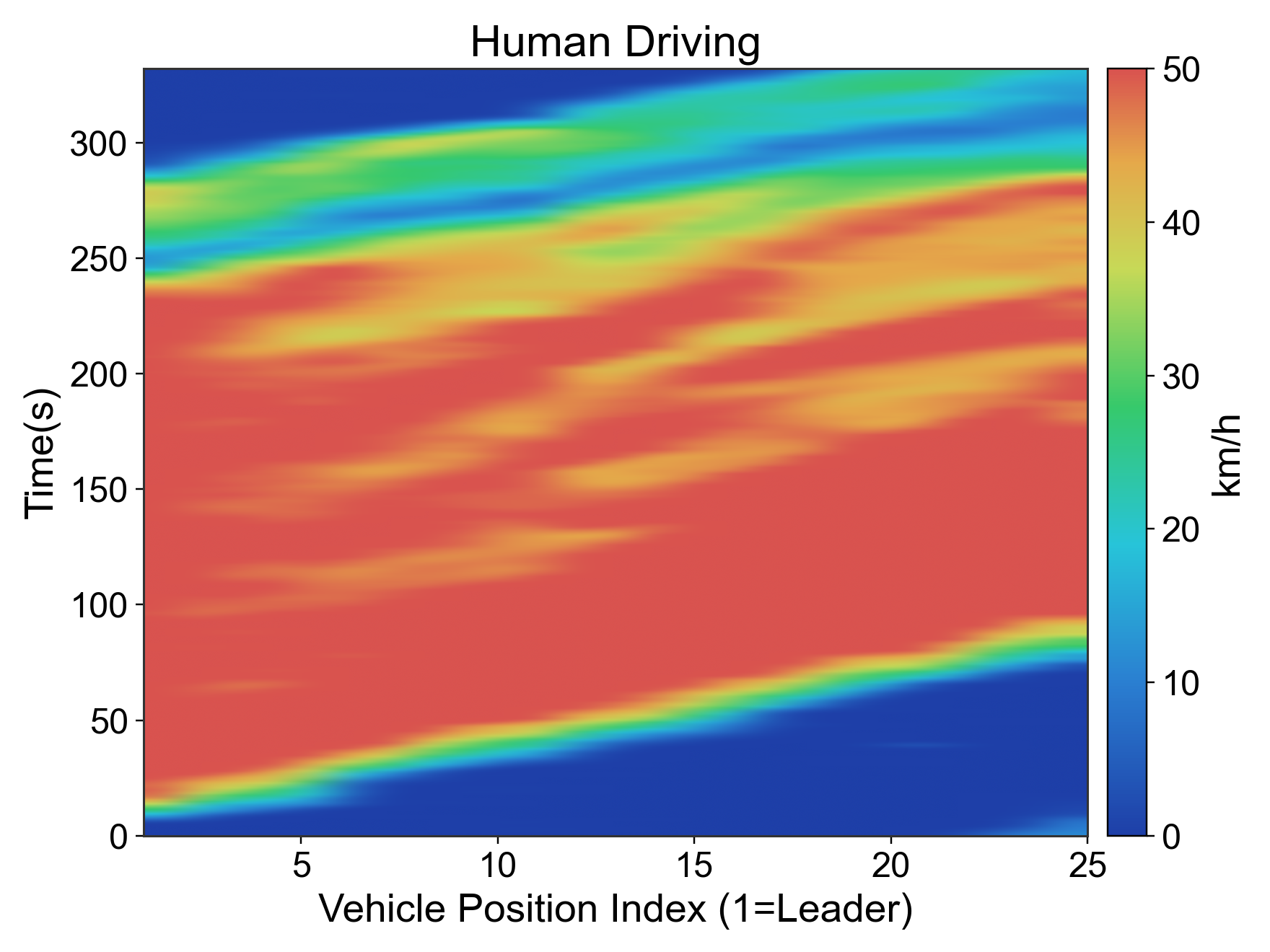}
(c)\includegraphics[width=0.47\linewidth]{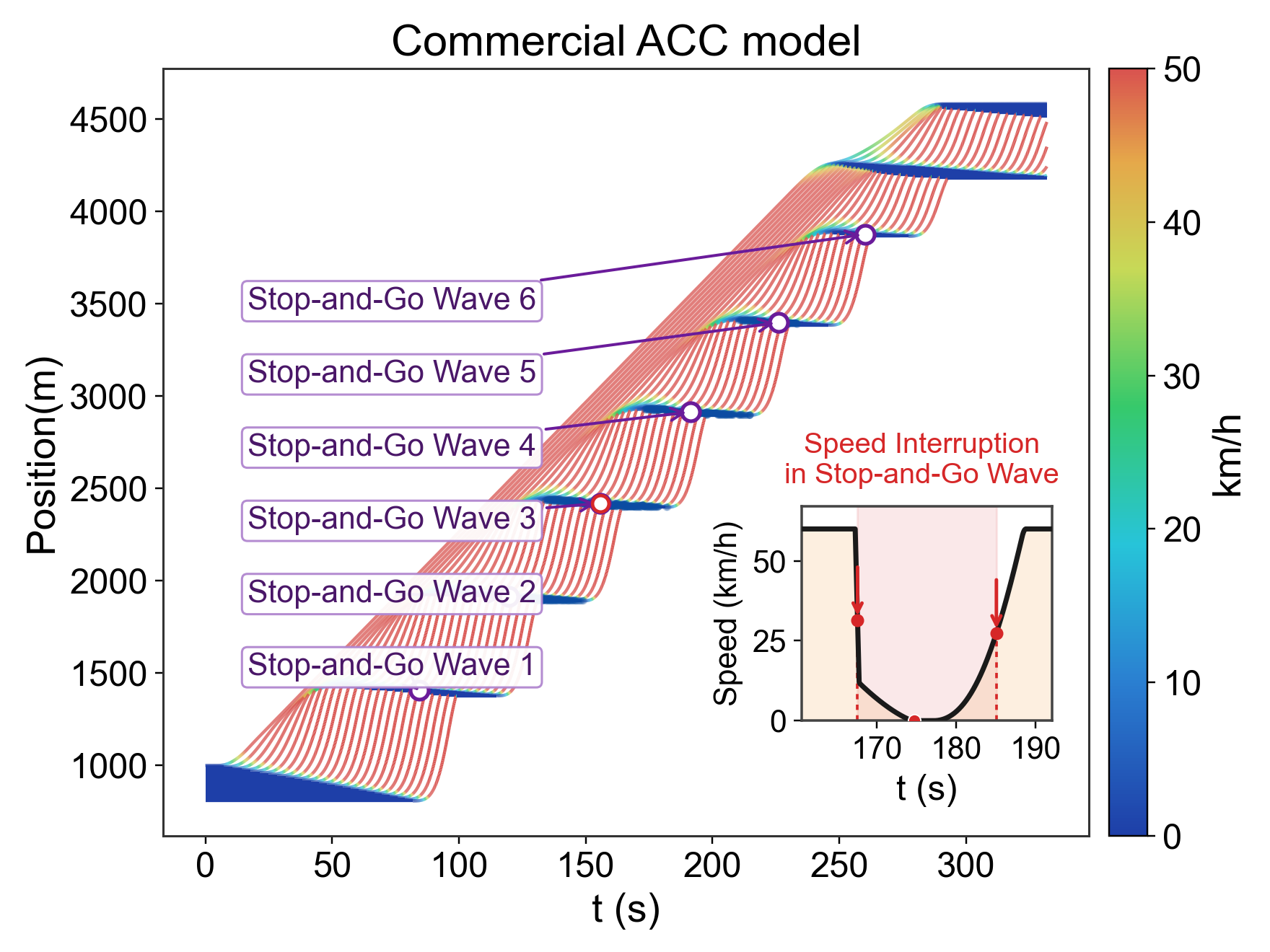}
(d)\includegraphics[width=0.47\linewidth]{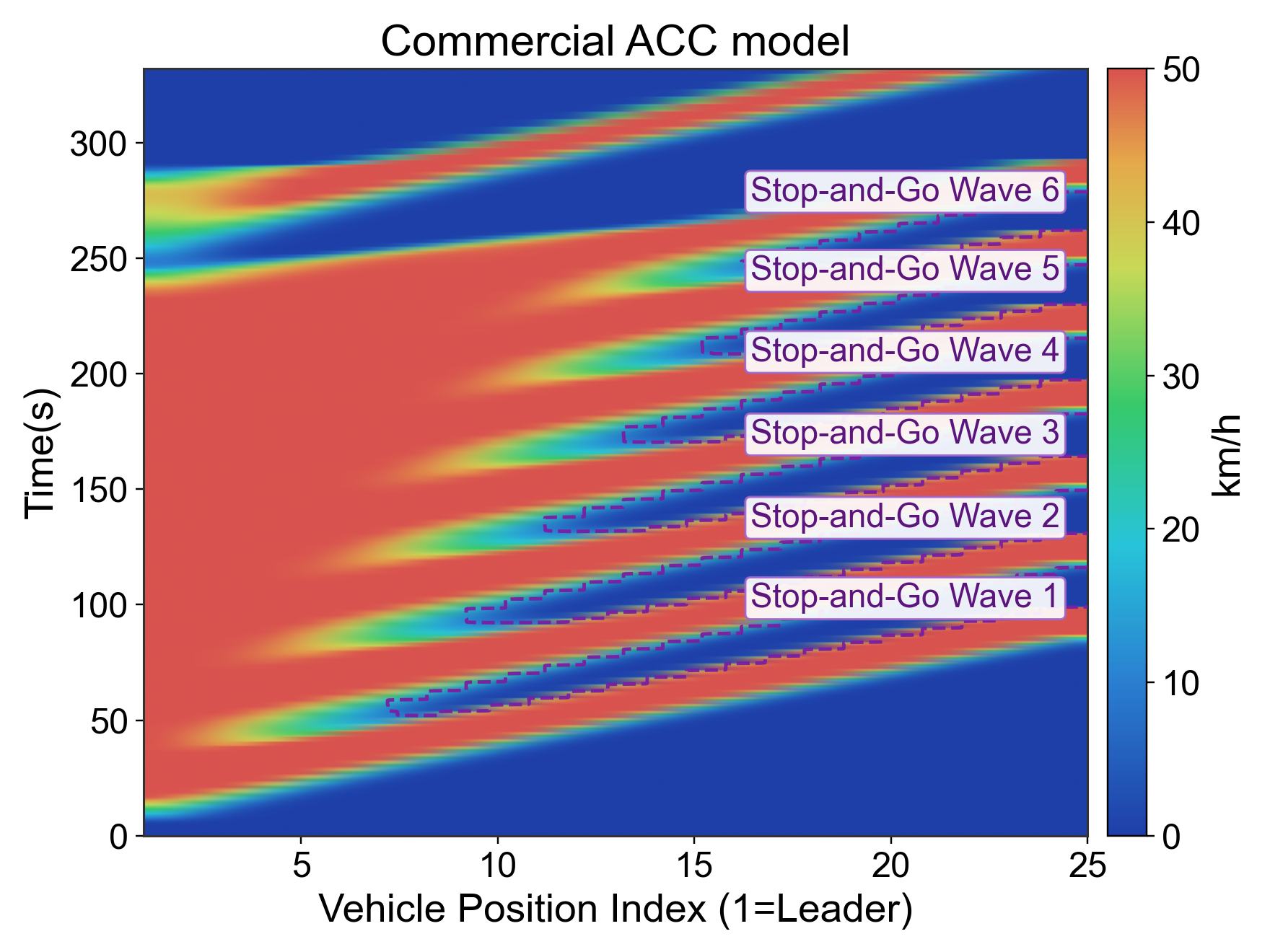}
\caption{\textbf{Human platoons dissipate perturbations; an ACC-model platoon amplifies them into stop-and-go waves.} \textbf{(a)} Empirical space--time trajectories from the 25-vehicle human-driven platoon (Hefei experiment, Run~14, $v_{\mathrm{lead,mean}} \approx 39$~km~h$^{-1}$; see Supplementary Section~S1 for run-selection criteria). Individual fluctuations remain incoherent and diffuse, allowing the collective system to attenuate perturbations and maintain smooth space--time evolution. The inset confirms a wide window of stable operation. \textbf{(b)} Spatiotemporal speed field for the human platoon. Speed variations stay fragmented and locally dissipated, with no persistent wave structure. \textbf{(c)} Microscopic space--time trajectories simulated via an empirically calibrated commercial ACC model~\cite{Gunter2021,gunterModelBasedStringStabilityAdaptive2020}. Small variations in leader motion trigger self-sustaining stop-and-go oscillations that cascade through successive followers. The inset reveals the abrupt velocity collapse within each oscillation cycle. \textbf{(d)} Spatiotemporal speed field for the ACC-model platoon. Homogeneous rule-based responses generate six distinct, highly organised low-speed bands that propagate backwards and intensify, indicating amplification rather than attenuation.}
\label{fig:2}
\end{figure*}

\section*{Results}

We systematically study the behavioural divergence between human adaptive flexibility and algorithmic rigidity across three complementary, interconnected scales of observation. We draw on four datasets: D1 (Hefei 25-vehicle platoon experiment), D2 (controlled 11-driver protocol), D3 (publicly available NGSIM highway database), and D4 (OpenACC repository of 22 production ACC systems), supplemented by an empirically calibrated commercial-ACC simulation. First, to evaluate emergent macro-scale behaviour, we contrast D1 against controlled multi-agent simulations governed by the commercially calibrated ACC model reported by Gunter \emph{et al.} (Fig.~\ref{fig:2}). Second, to identify the microscopic rules driving these collective outcomes, we map the steady-state speed--spacing boundaries of human operators estimated independently from D1--D3 against those of the ACC data (Fig.~\ref{fig:3}). Third, to unmask the foundational behavioural mechanism responsible for the observed human spacing paradigm, we analyse D2---a controlled protocol in which 11 drivers each completed multiple repeated runs under strictly homogeneous input conditions (109 runs in total)---to characterise the velocity-dependent structure of time-headway distributions (Fig.~\ref{fig:4}). This multi-scale progression establishes a rigorous empirical bridge, tracing a direct path from emergent macroscopic phenomena down to microscopic structural boundaries, and ultimately, to the underlying statistical and behavioural mechanics of human adaptability.

\subsection*{Automation produces coherent traffic oscillations whereas human driving dissipates them}

Traditional theories of car-following predict that human driving should be string-unstable: reaction time delays introduce phase lag between successive vehicles~\cite{Wilson2011LinearStability}, stochastic spacing fluctuations compound as they pass downstream~\cite{Jiang2018}, and drivers' limited look-ahead---typically restricted to the immediate leader---prevents anticipatory deceleration~\cite{Stern2018DissipationWaves}. Under these conditions, even minor speed fluctuations from the leader are expected to grow in amplitude through the platoon, eventually organising into persistent stop-and-go waves visible in space--time trajectories and speed heat maps. Yet despite this theoretical consensus, direct empirical tests using large, well-controlled platoon experiments have been rare.

We examined this directly using the Hefei 25-vehicle platoon experiment (D1; see Methods). Contrary to the traditional expectation, no congestion wave emerged: individual fluctuations remained incoherent and spatially diffuse, trajectories evolved smoothly across all 25 vehicles (Fig.~\ref{fig:2}a; the inset confirms that speed remained bounded within a stable operating window throughout the run), and the speed field showed no organised low-speed bands (Fig.~\ref{fig:2}b)---a property we term \emph{collective stability}.

One might then assume that a commercially implemented ACC system, applying deterministic rules consistently across all vehicles, would perform at least as well. To make a direct comparison---since no large-scale naturalistic ACC platoon dataset exists---we simulated a commercially calibrated ACC model using parameters from Gunter et al.~\cite{Gunter2021,gunterModelBasedStringStabilityAdaptive2020} (see Methods for details). Strikingly, the result is the opposite: small variations in leader motion triggered self-sustaining stop-and-go oscillations that cascaded through successive followers (Fig.~\ref{fig:2}c; the inset reveals an abrupt velocity collapse within each oscillation cycle), and six distinct low-speed bands developed and intensified over the run (Fig.~\ref{fig:2}d)---the very phenomenon that the human platoon suppressed. This raises an immediate question: what microscopic mechanism enables human drivers to achieve collective stability?

\subsection*{Adaptive segmentation in the speed--spacing relation reveals a robust microscopic turning point in human driving}

\begin{figure*}[htbp]
\centering
(a)\includegraphics[width=0.47\linewidth]{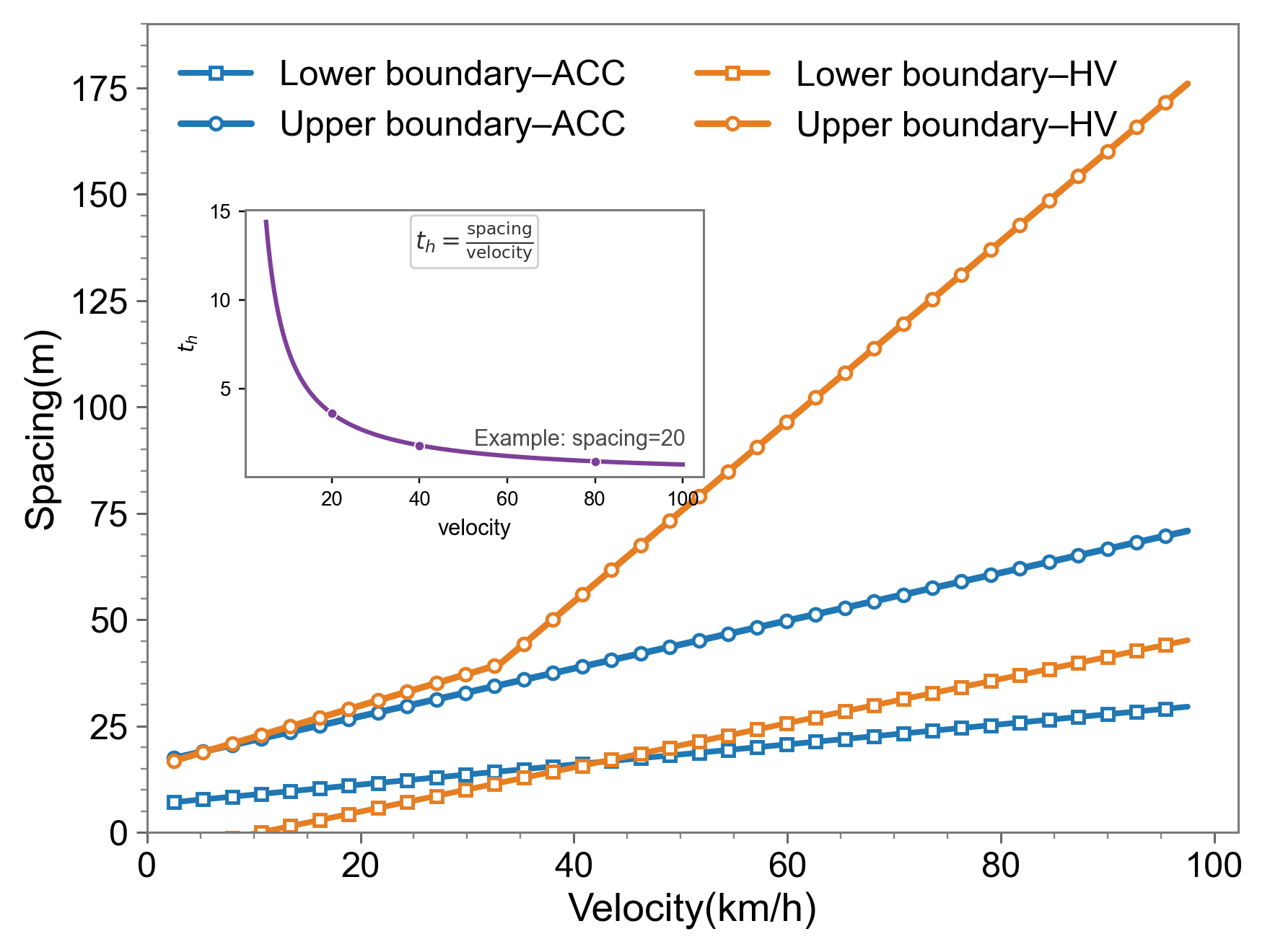}
(b)\includegraphics[width=0.47\linewidth]{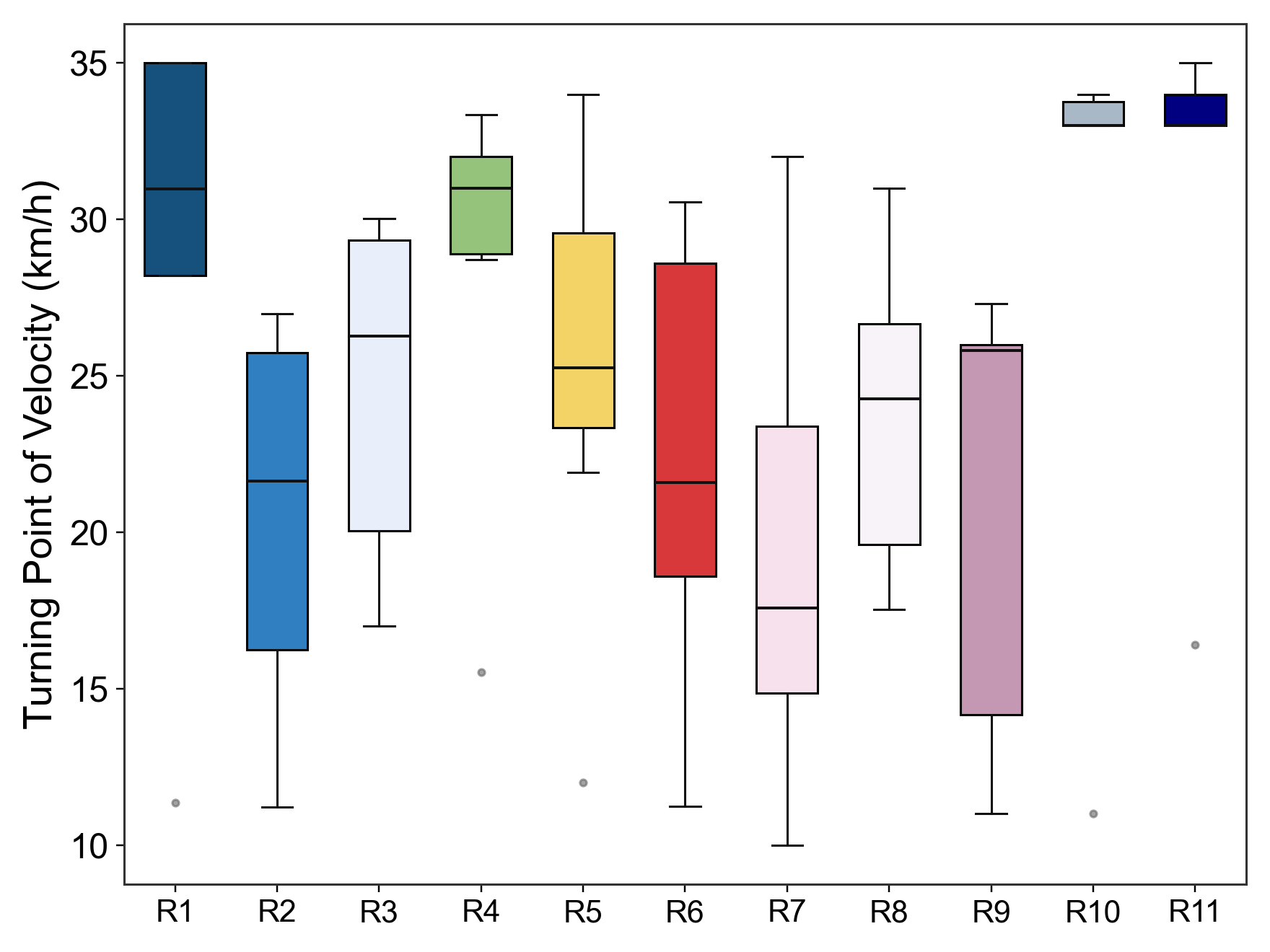}
\caption{\textbf{Microscopic turning in the speed--spacing relation is a robust feature of human driving across independent datasets.} \textbf{(a)} Empirical lower and upper bounds of the speed--spacing relation for human-driven vehicles and commercially implemented ACC. Under human driving, the upper boundary exhibits a clear turning point: spacing increases gradually at low speeds but rises more sharply beyond a critical speed. Under ACC, both boundaries increase approximately linearly with speed, consistent with a near-constant time-headway rule. This upper-boundary turning in human driving is consistent across all three human-driving datasets (D1, D2, and D3); see Supplementary Section~S2.7 for dataset-specific results. The inset illustrates the interpretation of time headway. Note: in the speed--spacing relation $s \approx s_0 + vT$, the follower speed $v$ is expressed in m~s$^{-1}$ for dimensional consistency, even though figure axes display speed in km~h$^{-1}$. \textbf{(b)} Distribution of turning-point speeds estimated for each of the 11 drivers in D2 (D2 comprised 109 repeated runs in total; labels R1--R11 identify individual drivers, each contributing multiple repeated runs under identical external conditions). The turning point is consistently present across drivers, confirming it as a robust property of human speed--spacing behaviour rather than an artefact of a single experimental trial (see Supplementary Section~S2.7 for further validation).}
\label{fig:3}
\end{figure*}

To answer this question, we study the microdynamics of driving behaviour. Specifically, we analyse how car followers regulate spacing as speed changes. In car-following, spacing (the front-bumper-to-rear-bumper distance between a following vehicle and its immediate leader, denoted $s$) can be approximated as $s \approx s_0 + vT$, where $s_0$ is the minimum jam spacing and $T$ is the time headway (the elapsed time between two consecutive vehicles passing a fixed reference point); as velocity $v$ increases, drivers tend to widen their following distance to maintain safety~\cite{Jiang2014NatureCarFollowing,Jiang2015,Tian2019SpeedAdaptation}. The speed--spacing relation captures this behaviour at the microscopic level; its lower and upper boundaries---estimated as the 5th and 95th percentiles of spacing across speed bins~\cite{Jabari2012StochasticTraffic}---define the envelope within which typical following observations lie, and together characterise how aggressively or cautiously a driver regulates space. In Fig.~\ref{fig:3}, we compare the speed--spacing relation for human drivers and commercially implemented ACC. The inset in Fig.~\ref{fig:3}a illustrates the interpretation of time headway: for a fixed spacing, $T = s/v$ decreases hyperbolically as velocity increases, meaning that sustaining a larger gap at higher speed is equivalent to maintaining a larger time headway. Since spacing depends linearly on $T$, the shape of the speed--spacing envelope is determined by how $T$ varies with speed, and is therefore shaped by the full headway distribution, not merely its mean.

As speed increases, the fundamental requirement of safe following dictates that spacing should grow with velocity; both lower and upper boundaries are therefore expected to increase. Under human driving, the lower boundary remains roughly linear, but the upper boundary exhibits a distinct turning point: spacing increases moderately at low speeds and then rises more sharply beyond a critical velocity (Fig.~\ref{fig:3}a). This turning is robust across runs and drivers (Fig.~\ref{fig:3}b). By contrast, for commercially implemented ACC (D4), both lower and upper spacing boundaries increase approximately linearly across the observed speed range, consistent with a near-constant time-headway rule, and no turning point is present.

This turning corresponds to a change in the slope of the upper envelope of the speed--spacing relation. It indicates that human following does not obey a single proportional spacing rule across all speeds \cite{tianCarFollowingBehavioralStochasticity2021,Jiang2015,tianCellularAutomatonModelDynamical2017}. Instead, the growth of spacing with speed is segmented into at least two regimes. At lower speeds, spacing expands gradually; at higher speeds, the expansion becomes substantially steeper. This pattern is absent in D4, whose upper boundary retains a nearly uniform slope.

The significance of this structural contrast lies in the stability mechanism it implies. Beyond the critical speed, human spacing grows disproportionately fast, maintaining a buffer that exceeds what a constant-headway rule would prescribe. When a small perturbation---such as a slight deceleration by the leader---arrives, this excess gap is consumed gradually rather than requiring an abrupt braking response, damping the disturbance before it reaches the next follower. Repeated across successive vehicles in a platoon, this adaptive buffering progressively weakens disturbances rather than allowing them to accumulate into coherent waves. Commercially implemented ACC, applying a near-constant time-headway rule, provides no such disproportionate buffer: any deceleration triggers an immediate proportional response, transmitting the full perturbation downstream and allowing it to amplify---consistent with the stop-and-go wave patterns observed in Fig.~\ref{fig:2}.

The turning point is also robust. Figure~\ref{fig:3}b shows the distribution of turning-point speeds estimated across repeated experimental rounds. Although the precise turning speed varies across runs---spanning roughly 14--34~km~h$^{-1}$ across the 11 repeated runs (R1--R11)---the breakpoints repeatedly appear rather than arising as isolated outliers. This variability is expected: human car-following behaviour is idiosyncratic, and the exact speed at which a driver transitions from compact to expanded spacing can depend on personal thresholds, vehicle dynamics, and road context. The key observational claim is therefore not that all drivers share a single universal critical speed, but that a critical-speed-like structural nonlinearity is consistently present across human drivers and datasets (Supplementary Section~S2.7), while being consistently absent in commercial ACC.

This observation is consistent with prior work showing that human time-headway selection adapts systematically with speed~\cite{Tian2019SpeedAdaptation,Jiang2015,tianCarFollowingBehavioralStochasticity2021}, and that perceived headway criticality is itself velocity-dependent~\cite{Dey2009DesiredTimeGap}. These findings suggest that a structural change in spacing behaviour around a critical speed is a plausible and expected feature of human driving. In our system, the repeated emergence of such a turning point indicates that this nonlinearity may be a key microscopic mechanism through which individual human-following behaviour contributes to the stability of the traffic system.

\subsection*{Adaptive variability in time headways explains the segmented upper boundary}

\begin{figure*}[htbp]
\centering
(a)\includegraphics[width=0.47\linewidth]{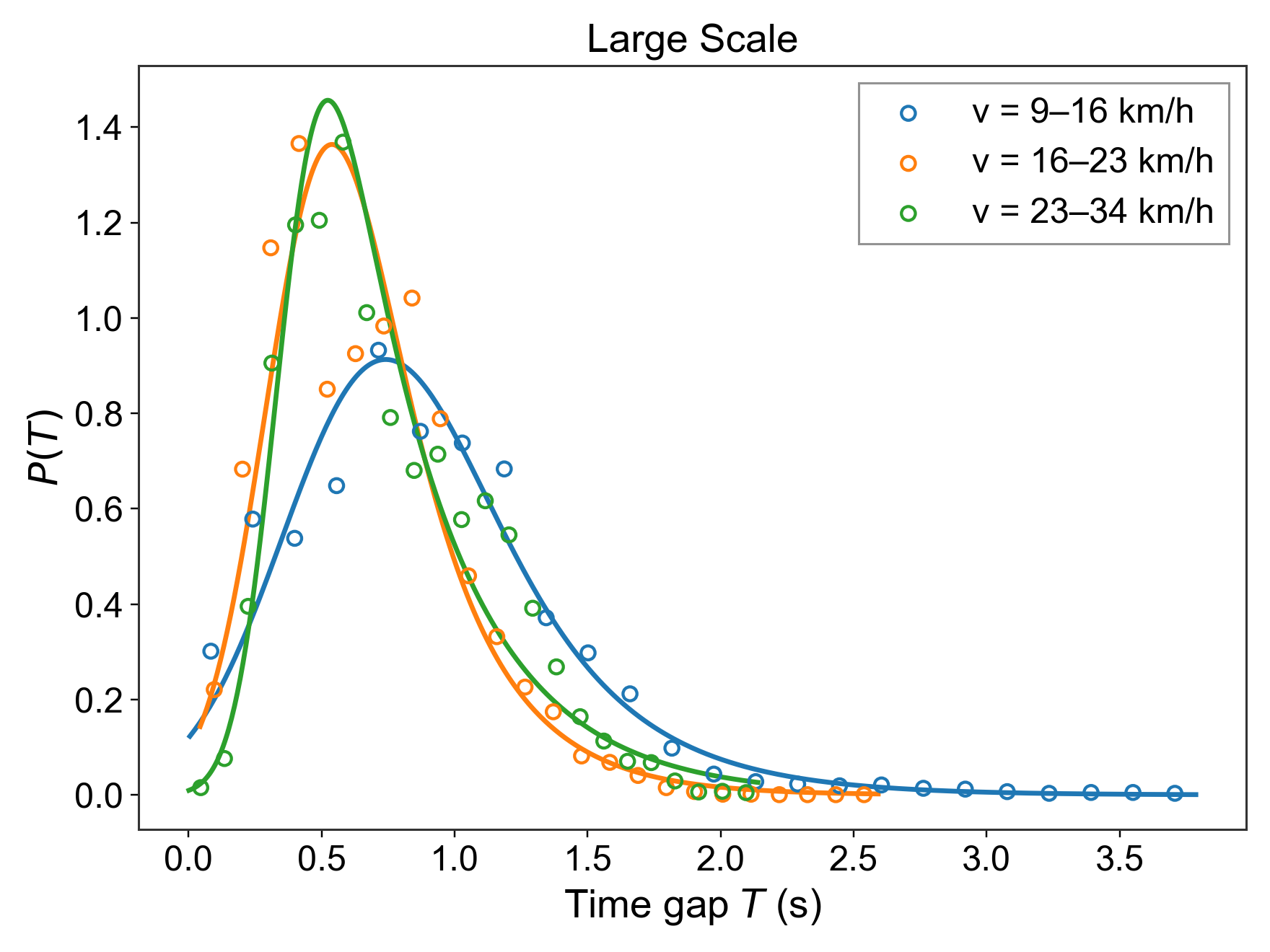}
(b)\includegraphics[width=0.47\linewidth]{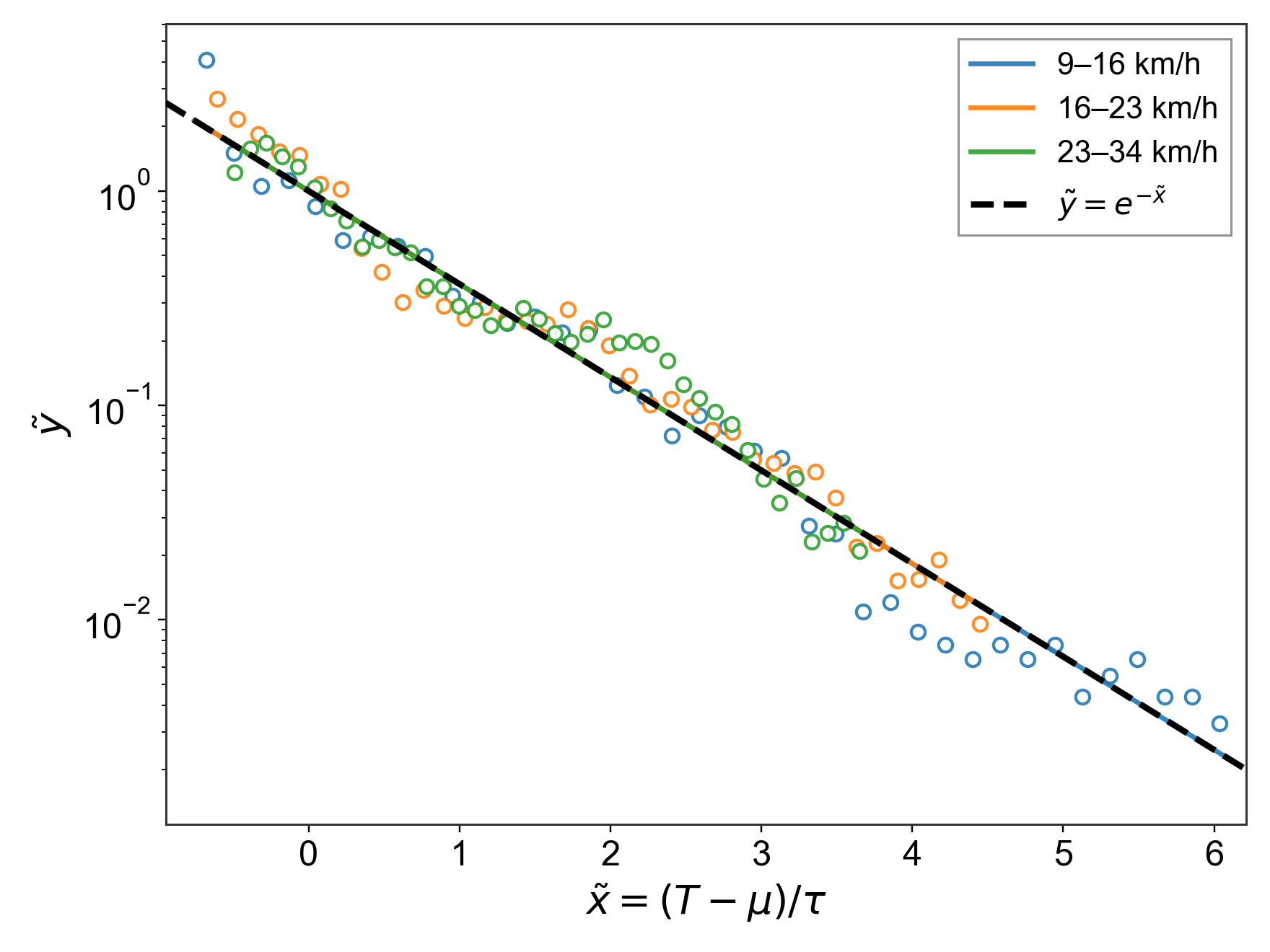}\\
(c)\includegraphics[width=0.95\linewidth]{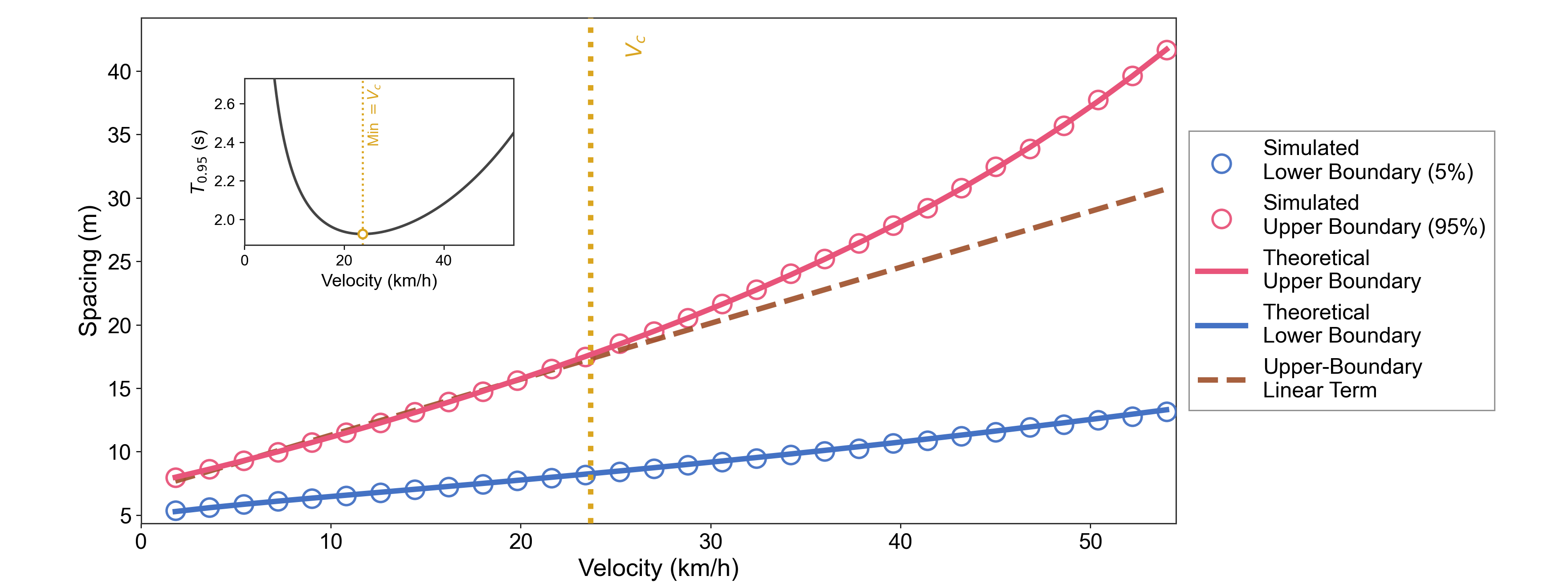}
\caption{\textbf{Ex-Gaussian time-headway statistics explain the segmented speed--spacing boundary in human driving.} \textbf{(a)} Empirical time-headway distributions for three representative speed bins from D2 (9--16, 16--23, and 23--34~km~h$^{-1}$). Hollow circles are empirical bin counts; solid curves are the corresponding maximum-likelihood ex-Gaussian fits~\cite{Lacouture2008ExGaussian,Branston1976HeadwayModels,Dey2009DesiredTimeGap}. Across all three speed regimes the ex-Gaussian form closely captures both the bulk of the distribution and the heavy right tail. \textbf{(b)} Collapse diagnostic on a logarithmic scale for the same speed bins (see Methods and Supplementary Section~S3 for the rescaling transformation). When plotted against the normalised coordinate $\tilde{x}=(T-\mu)/\tau$, the rescaled densities from all speed bins collapse onto the reference curve $e^{-\tilde{x}}$ (dashed black), confirming that the exponential-tail structure of the ex-Gaussian is preserved across the full velocity range. \textbf{(c)} Theoretical lower and upper speed--spacing boundaries derived by substituting the fitted ex-Gaussian headway distribution into $s \approx s_0 + vT$, compared with the corresponding numerically computed boundaries obtained via Monte Carlo sampling from the same fitted ex-Gaussian distributions. The close agreement validates the analytical approximations and shows that the observed upper-boundary bending can be reproduced directly from the statistical structure of human time headways. The inset shows the non-monotonic variation of $T_{0.95}(v)$, with a single minimum at the critical speed $V_c \approx 23.7~\mathrm{km~h^{-1}}$; below $V_c$ the upper boundary rises gradually, above it the boundary steepens markedly as the exponential tail grows disproportionately with speed. Together, these results suggest that the segmented spacing pattern arises from a continuous, speed-dependent reshaping of the time-headway distribution, rather than from a discrete switch between driving modes.}
\label{fig:4}
\end{figure*}

The turning in the human upper boundary provides a plausible microscopic link between adaptive following behaviour and collective stability; its statistical origin, however, remains to be explained. We therefore further analysed the distribution of time headways and its speed-dependence using D2 (Fig.~\ref{fig:4}). Prior studies of headway distributions have often adopted simplified parametric forms---most commonly Gaussian or log-normal models---to characterise following variability~\cite{Branston1976HeadwayModels,Dey2009DesiredTimeGap}. Whether these assumptions hold across the full speed range, and how the headway distribution evolves with velocity, has seldom been directly tested using naturalistic trajectory data. The controlled 11-driver experiment (D2) provides exactly this opportunity: repeated individual observations across a wide speed range allow direct characterisation of the empirical distribution of $T$ at each speed bin. As shown in Fig.~\ref{fig:4}a, the empirical time-headway distributions across all speed bins are well described by the ex-Gaussian form.

The ex-Gaussian random variable decomposes as $T = G + E$, where $G \sim \mathcal{N}(\mu,\sigma^2)$ captures the Gaussian core of typical, near-automatic following responses, and $E \sim \mathrm{Exp}(\lambda)$ with mean $\tau = 1/\lambda$ represents an exponential tail of occasional conservative gaps maintained for safety~\cite{Lacouture2008ExGaussian}. Both components are present across the observed speed range, indicating that human time headways encode a mixture of habitual proximity maintenance and deliberate safety-margin expansion. To further validate the ex-Gaussian characterisation, especially the exponential-tail behaviour across different speed regimes, we estimated the three ex-Gaussian parameters for each speed regime and applied a corresponding mathematical transformation to rescale the distributions (see Methods and Supplementary Section~S3). Under the ex-Gaussian assumption, these transformed distributions are expected to collapse onto a single exponential curve. As shown in Fig.~\ref{fig:4}b, the empirical distributions exhibit this collapse, providing further support for the ex-Gaussian model.

Crucially, this statistical structure deforms with speed: as velocity increases, the exponential component of the headway distribution also changes (see Methods and Supplementary Information). Furthermore, this speed-dependent reshaping enables a mathematical framework to analytically derive the upper and lower boundaries of the speed--spacing relationship (see Methods and Supplementary Section~S3). The lower boundary is well approximated by a linear function---consistent with the empirical findings---reflecting a near-proportional minimum safe spacing across speeds. The upper boundary is more complex: the framework predicts the existence of a critical speed $V_c \approx 23.7$~km~h$^{-1}$, obtained from $\mathrm{d}T_{0.95}/\mathrm{d}v = 0$ and consistent with the empirically observed turning range (see inset, Fig.~\ref{fig:4}c). Below $V_c$, the upper boundary rises gradually as the exponential tail of the headway distribution expands only moderately with speed, maintaining relatively compact safety margins. Above $V_c$, the boundary steepens markedly as the exponential tail grows disproportionately with speed, driving increasingly cautious spacing---a statistical signature of adaptive variability in human following behaviour. These analytical predictions are confirmed by Monte Carlo simulation from the same fitted ex-Gaussian distributions (Fig.~\ref{fig:4}c).

These results imply that the segmented upper boundary is a statistical signature of adaptive variability. Rather than following a fixed proportional rule, human drivers adjust their headway distribution across speed regimes, leading to a nonlinear expansion of spacing as conditions change. At the collective level, this variability provides a plausible mechanism for damping perturbations before they organise into coherent oscillations. By contrast, commercially implemented ACC exhibits a near-linear spacing rule and lacks this speed-dependent reshaping, which helps explain why disturbances persist and amplify in the ACC-like platoon.

\subsection*{Traffic stability, fuel consumption, and emissions across scenarios}

\begin{table*}[!ht]
\centering
\caption{Comparison of traffic performance and energy--emission metrics between human-driven vehicles (HV) and commercial ACC vehicles across different scenarios. \textit{Mean speed}: fleet-average speed (km~h$^{-1}$). \textit{Speed std}: standard deviation of instantaneous speed (km~h$^{-1}$). \textit{Acc RMS}: root-mean-square of vehicle acceleration (m~s$^{-2}$). \textit{Near-stop ratio}: fraction of time steps with speed below 5~km~h$^{-1}$ (dimensionless). \textit{Fuel/veh-km}: fuel consumption per vehicle per kilometre (L~veh$^{-1}$~km$^{-1}$). \textit{CO$_2$/veh-km}: CO$_2$ emissions per vehicle per kilometre (kg~veh$^{-1}$~km$^{-1}$). \textit{$\Delta$}: relative increase of the commercial ACC fleet over HV (\%); since CO$_2 = \text{Fuel}\times EF_{\mathrm{CO_2}}$ with constant $EF_{\mathrm{CO_2}}$, the relative change is identical for both quantities and a single column is reported. All $\Delta$ values are computed from unrounded intermediate results and may differ slightly from values derived from the rounded figures shown in the table. All metrics were computed over the 24 following vehicles, excluding the leader and the initial 40~s warm-up period. \textit{Acc RMS note}: the ACC simulation bounds the commanded longitudinal acceleration to $\pm3$~m~s$^{-2}$ (consistent with commercial ACC comfort limits); the Acc~RMS values reported here include both bounded commanded accelerations and hard-braking contributions from stop-and-go events. Fuel and CO$_2$ estimates are computed from tractive power only (positive acceleration terms), so they reflect the bounded commanded accelerations.}
\label{tab:scenario_comparison}
\resizebox{\textwidth}{!}{
\begin{tabular}{llccccccc}
\toprule
\rowcolor{headergray}
Scenario & Fleet & \makecell[c]{Mean speed\\(km h$^{-1}$)} & \makecell[c]{Speed std\\(km h$^{-1}$)} & \makecell[c]{Acc RMS\\(m s$^{-2}$)} & \makecell[c]{Near-stop\\ratio} & \makecell[c]{Fuel/veh-km\\(L veh$^{-1}$ km$^{-1}$)} & \makecell[c]{CO$_2$/veh-km\\(kg veh$^{-1}$ km$^{-1}$)} & $\Delta$ (\%) \\
\midrule
Stable & HV  & 28.149 & 5.929  & 0.330 & 0.003 & 0.061 & 0.140 & -- \\
\rowcolor{rowgray}
Stable & ACC & 28.381 & 31.356 & 6.188 & 0.355 & 0.285 & 0.659 & +370.4 \\

Mild perturbation & HV  & 28.663 & 11.620 & 0.349 & 0.086 & 0.063 & 0.145 & -- \\
\rowcolor{rowgray}
Mild perturbation & ACC & 29.596 & 35.065 & 6.688 & 0.405 & 0.311 & 0.718 & +395.5 \\

Strong perturbation & HV  & 44.545 & 26.794 & 0.468 & 0.152 & 0.074 & 0.172 & -- \\
\rowcolor{rowgray}
Strong perturbation & ACC & 44.659 & 47.305 & 8.482 & 0.360 & 0.320 & 0.739 & +330.7 \\

Repeated stop-and-go & HV  & 12.927 & 7.187  & 0.309 & 0.102 & 0.082 & 0.190 & -- \\
\rowcolor{rowgray}
Repeated stop-and-go & ACC & 13.367 & 18.506 & 3.157 & 0.438 & 0.221 & 0.510 & +168.5 \\
\bottomrule
\end{tabular}
}
\end{table*}

Table~\ref{tab:scenario_comparison} shows that, under the modelled scenarios and the calibrated ACC parameters described in Methods, the commercial ACC fleet consistently exhibited larger traffic oscillations and substantially worse energy--emission performance than the human-driven fleet (HV) across all four scenarios, despite having similar mean speeds. These fuel and emissions figures are scenario-specific model outcomes derived from a single calibrated ACC system; they are not intended as universal predictions for all commercial ACC implementations. In the stable scenario, the mean speeds of the two fleets were nearly identical (28.1~km~h$^{-1}$ for HV versus 28.4~km~h$^{-1}$ for ACC), but the commercial ACC fleet showed a much larger speed standard deviation (31.4 versus 5.9~km~h$^{-1}$), acceleration RMS (6.19 versus 0.33~m~s$^{-2}$), and near-stop ratio (0.355 versus 0.003). This stronger oscillatory behaviour was accompanied by markedly higher fuel consumption and CO$_2$ emissions, with fuel use increasing from 0.061 to 0.285~L~veh$^{-1}$~km$^{-1}$ and CO$_2$ emissions rising from 0.140 to 0.659~kg~veh$^{-1}$~km$^{-1}$, corresponding to an increase of $+$370.4\% in both fuel consumption and CO$_2$ emissions.

A similar pattern was observed under mild perturbation. Although the commercial ACC fleet maintained a slightly higher mean speed than the HV fleet (29.6 versus 28.7~km~h$^{-1}$), it exhibited much larger fluctuations, with the speed standard deviation increasing from 11.6 to 35.1~km~h$^{-1}$ and the acceleration RMS increasing from 0.35 to 6.69~m~s$^{-2}$. The near-stop ratio also rose sharply, from 0.086 in HV to 0.405 in ACC. These amplified oscillations translated into a $+$395.5\% increase in both fuel consumption and CO$_2$ emissions relative to the HV baseline, reaching 0.311~L~veh$^{-1}$~km$^{-1}$ (fuel) and 0.718~kg~veh$^{-1}$~km$^{-1}$ (CO$_2$).

Under strong perturbation, the contrast remained pronounced. The two fleets had comparable mean speeds (44.5~km~h$^{-1}$ for HV and 44.7~km~h$^{-1}$ for ACC), but the commercial ACC fleet exhibited substantially stronger instability, with the speed standard deviation increasing from 26.8 to 47.3~km~h$^{-1}$ and the acceleration RMS rising from 0.47 to 8.48~m~s$^{-2}$. The near-stop ratio also remained much higher in the ACC fleet (0.360 versus 0.152). In parallel, fuel consumption increased from 0.074 to 0.320~L~veh$^{-1}$~km$^{-1}$ and CO$_2$ emissions from 0.172 to 0.739~kg~veh$^{-1}$~km$^{-1}$, representing an increase of $+$330.7\% in both fuel consumption and CO$_2$ emissions.

Even in the repeated stop-and-go scenario, where both fleets operated at relatively low mean speeds, the commercial ACC fleet still exhibited substantially greater fluctuations and lower efficiency. Relative to HV, the ACC fleet had a higher speed standard deviation (18.5 versus 7.2~km~h$^{-1}$), a much larger acceleration RMS (3.16 versus 0.31~m~s$^{-2}$), and a markedly elevated near-stop ratio (0.438 versus 0.102). Fuel consumption increased from 0.082 to 0.221~L~veh$^{-1}$~km$^{-1}$ and CO$_2$ emissions from 0.190 to 0.510~kg~veh$^{-1}$~km$^{-1}$, corresponding to an increase of $+$168.5\% in both fuel consumption and CO$_2$ emissions.

Taken together, these results show that the degradation of the commercial ACC fleet is not primarily associated with reduced mean speed, but with amplified temporal fluctuations in speed and acceleration. Across all scenarios, the ACC fleet experienced more frequent near-stop events, greater acceleration variability, and substantially higher fuel and emissions costs. This pattern suggests that the main collective disadvantage of the commercial ACC fleet lies in its tendency to amplify perturbations into stop-and-go-like oscillations, thereby imposing a large energetic and environmental penalty even when average travel speed remains similar to that of the HV fleet~\cite{He2020EnergyImpact,Apostolakis2023}.

\section*{Discussion}

This study shows that the sharp contrast between human and automated network stability is rooted in behaviour rather than computation, drawing on real human-driving data, a commercial ACC benchmark, and a calibrated ACC model simulation. Our findings reveal that current rule-based longitudinal control increases collective oscillations compared to human platoons---not because of differences in sensing or actuation precision, but because of how following behaviour is organised: human drivers regulate spacing variably and adaptively, whereas commercial ACC relies on rigid, near-proportional feedback. The resulting platoon dynamics are consistent with large-scale experimental findings showing that oscillation growth in traffic follows reproducible statistical laws rather than random, trial-specific anomalies~\cite{Jiang2018}.

Three connected lines of evidence support this paradigm shift. First, at the macroscopic level, the homogeneous ACC-model platoon spontaneously generates stop-and-go waves from small disturbances, whereas the human platoon disperses these perturbations, keeping speed fluctuations diffuse and fragmented. Second, at the microscopic level, human drivers exhibit a non-monotonic spacing strategy with a segmented upper boundary and a clear structural turning point across datasets---markedly different from commercial ACC, which maintains a fixed, linear time-headway policy. This turning structure appears consistently across D1, D2, and D3, confirming that it is a robust property of human driving rather than an artefact specific to the present experiments. Third, this turning pattern is explained by a continuous, velocity-dependent reshaping of the time-headway distribution, with the rising safety boundary arising from the skewed tail of headway variability rather than from a rule-based switch between driving modes.

Together, these empirical and theoretical findings show that human behavioural variability is not merely sub-optimal noise or cognitive weakness---individual adaptive irregularity plays a significant role in collective regulation. By continuously reshaping spacing responses across speed regimes, human drivers introduce a natural nonlinear damping mechanism that absorbs local errors before they organise into coherent oscillations. This interpretation builds on the complex systems and traffic physics literature, in which oscillation growth in traffic is linked not only to linear instability but also to behavioural thresholds, local randomness, and state-based response rules~\cite{Laval2014ParsimoniousOscillation,Tian2019SpeedAdaptation,Bouadi2022StochasticStringStability}. Speed-dependent distribution reshaping constitutes an adaptive damping barrier at the behavioural level, distinct from random white noise~\cite{Treiber2018,Chen2012}. By contrast, fixed proportional spacing compels automated followers to react nearly identically to disturbances, a uniformity that facilitates the phase-locking of local errors into collective stop-and-go waves. True network stability emerges not from eliminating variability, but from deploying the right kind of adaptive variability.

This interpretation clearly defines the limits of our conclusions. Our results do not claim that automation is always destabilising, or that human driving is flawless. The automated benchmark comes from current commercial ACC data and a rule-based model calibrated with field data. The limitation we identify concerns a specific type of deterministic controller, not the whole field of future autonomous designs. More adaptive or learning-based AI systems may act differently. Still, our result provides a key lesson: simple commercial proportional control does not reproduce the emergent stability seen in human behaviour.

The main design implication is clear. Building resilient automation means going beyond traditional metrics like error minimisation. Highly precise but inflexible machine agents stay fragile at scale. Future automated controllers must use speed-dependent adaptation or state-based spacing. These allow adjustment as conditions change. The goal is not to mimic human inconsistency, but to recover the stabilising mechanics in human adaptive behaviour.

Several scientific and practical issues remain for future work. Our analysis focused on single-lane car-following. It does not show how damping generalises to multi-lane roads, mixed fleets, or situations with frequent lane changes. Our 11-driver experiment was designed to study headway deformation, not as a direct comparison to automation. Also, our macroscopic dynamics use a commercial ACC model calibrated to field data---not data from real-world autonomous fleets. These modelling strategies help isolate behavioural mechanisms, but they limit our insights. This approach matches reality: commercial ACCs are proprietary black boxes, so we cannot access their internal logic. Using calibrated proxies gives us reproducible, scenario-independent evaluations while staying tied to observed commercial behaviour. A fuller discussion of these limitations---including the scope of the automated benchmark, single-lane constraints, fleet composition assumptions, and fuel model generalisability---is provided in Supplementary Section~S6.

In conclusion, our results establish a rigorous bridge linking microscopic behavioural variability to macroscopic collective stability. Human behavioural groups exhibit superior macro-stability not because individual operators possess higher sensing accuracy, but precisely because their responses are more adaptively variable. This intrinsic individual flexibility continuously reshapes safety margins across speed regimes, systematically suppressing the destructive synchronisation that gives rise to stop-and-go waves. The broader, cross-disciplinary lesson is clear: if we are to build resilient automation and robust human-agent collectives, artificial systems must be engineered not merely for local tracking accuracy, but for emergent behavioural adaptability.

\section*{Methods}

\subsection*{Overview}

This study draws on four complementary datasets (D1--D4) whose properties and roles are detailed in the following subsection. In addition, to represent commercial ACC dynamics in controlled platoon simulations, we use a second-order delay-differential model with parameter settings calibrated from field data by Gunter~\emph{et~al.}~\cite{Gunter2021,gunterModelBasedStringStabilityAdaptive2020}.

\subsection*{Empirical data sources}

D1 is a large-scale platoon experiment in which 25 human-driven vehicles followed each other in a single file on a straight highway in Hefei, China~\cite{Jiang2015}. Their positions and speeds were continuously logged over multiple repeated runs, providing a direct empirical record of how speed disturbances propagate---or die out---through a long human convoy.

D2 is a controlled car-following experiment conducted at a closed test track in China. A lead vehicle executed a prescribed series of speed changes while 11 drivers each followed it on repeated runs under identical external conditions, yielding multiple observations of each driver responding to the same stimulus. This design isolates natural variation in individual following behaviour from differences in the driving environment.

D3 is the NGSIM naturalistic dataset, comprising vehicle trajectory recordings collected on US freeways (I-80 and US-101, California). Its large size and real-world diversity make it a valuable independent check on whether the following-distance patterns found in D1 reflect a general feature of human driving rather than a condition specific to the Hefei experiment.

D4 is the OpenACC database~\cite{Makridis2021}, an open collection of speed and following-distance recordings from 22 commercially equipped ACC vehicles gathered across five test campaigns in Europe. Its coverage of multiple vehicle models and road conditions provides an empirical picture of how production ACC systems regulate following distance in practice.

These datasets play distinct roles. D1 captures collective platoon behaviour (Fig.~2) and the human following-distance pattern across speeds (Fig.~3). D2 reveals how individuals distribute their time gaps at different speeds (Fig.~4). D3 provides an independent naturalistic validation of the human pattern, with results in Supplementary Section~S2.7 and Fig.~3a. D4 characterises the following-distance pattern of production ACC systems (Fig.~3a). To compare the collective behaviour observed in D1 against an ACC counterpart, we used D4 as empirical grounding to calibrate a commercial ACC simulation model, generating a matched platoon simulation as a direct comparison. Because D4 combines vehicles, headway settings, and road conditions across campaigns, it is not directly suitable for the collective comparison in Fig.~2 on its own; that comparison therefore uses the calibrated model described below, with D4 providing empirical grounding for the model's following-distance behaviour.

\subsection*{Trajectory processing and statistical boundary estimation}

From the trajectory data, instantaneous spacing, speed, and time headway were extracted for each follower--leader pair. Observations were filtered to retain only steady-state car-following segments in which the follower's speed closely tracked the leader's. To characterise how following distance varies with speed, the typical range of spacing was summarised by its lower (5th-percentile) and upper (95th-percentile) envelopes across speed levels; a piecewise linear fit was then used to detect any turning point in the upper envelope. Full details of the preprocessing protocol, filtering criteria, boundary estimation, and turning-point detection are provided in Supplementary Sections~S1 and~S2.

\subsection*{Adaptive-variability modelling and calibrated commercial ACC simulation}

Human time headways within each speed bin were modelled using an ex-Gaussian distribution, which decomposes into a Gaussian core (typical near-automatic following responses) and an exponential tail (occasional conservative gaps):
\begin{align}
f(T;\mu,\sigma,\lambda)
={}& \lambda
\exp\!\left[
\lambda\!\left(
\mu+\frac{\lambda\sigma^2}{2}-T
\right)
\right]
\nonumber\\
&\times
\Phi\!\left(
\frac{T-\mu-\lambda\sigma^2}{\sigma}
\right).
\end{align}
where $\mu$ and $\sigma$ describe the Gaussian component, $\lambda$ governs the exponential tail, and $\Phi$ is the standard normal CDF. Parameters were estimated by maximum-likelihood estimation for each speed bin.

To verify that the exponential-tail structure is preserved across speed regimes, we apply the collapse transformation
\begin{align}
\tilde{y}(T) = \frac{\tau\,f(T)\,\exp\!\left(\sigma^2/2\tau^2\right)}{\Phi\!\left[(T-\mu)/\sigma - \sigma/\tau\right]},
\end{align}
plotted against $\tilde{x} = (T-\mu)/\tau$. In the right tail (where the exponential component dominates), $\tilde{y}$ converges to $e^{-\tilde{x}}$ regardless of speed bin, providing a distribution-independent confirmation that the tail structure is preserved across velocities (Fig.~\ref{fig:4}b). Full derivation is given in Supplementary Section~S3.

Using the approximation $S=S_0+vT$, the 5th- and 95th-percentile spacing boundaries follow analytically from the fitted parameters. Define $\tau(v)=1/\lambda(v)$ and $z_{0.05}=\Phi^{-1}(0.05)\approx-1.645$. The lower boundary uses a Cornish--Fisher approximation around the Gaussian core of the ex-Gaussian distribution:
\begin{align}
T_{0.05}(v) \approx \mu + \tau(v) + z_{0.05}\sqrt{\sigma^2+\tau(v)^2}
+ \frac{\tau(v)^3\!\left(z_{0.05}^2-1\right)}{3\!\left[\sigma^2+\tau(v)^2\right]},
\end{align}
and the upper boundary uses an exponential right-tail approximation:
\begin{align}
T_{0.95}(v) \approx \mu + \frac{\sigma^2}{2\tau(v)} - \tau(v)\ln(0.05),
\end{align}
so that $S_{0.95}(v)=S_0+v\,T_{0.95}(v)$ and $S_{0.05}(v)=S_0+v\,T_{0.05}(v)$. Substituting $\lambda(v)=Av\,e^{-Bv}$, the upper-bound headway $T_{0.95}(v)$ reaches its minimum at a critical velocity
\begin{align}
V_c = \frac{1}{B},
\end{align}
obtained from $\mathrm{d}T_{0.95}/\mathrm{d}v=0$, which evaluates to approximately $23.7$~km~h$^{-1}$ for the calibrated parameters---consistent with the empirically observed turning range. These boundaries closely match Monte Carlo quantiles sampled from the same fitted distributions (Fig.~\ref{fig:4}c). Full derivations are given in Supplementary Section~S3.

To reproduce commercial ACC platoon dynamics in Fig.~2, we simulated automated car-following using a second-order delay-differential model calibrated from field data by Gunter \emph{et al.}~\cite{Gunter2021,gunterModelBasedStringStabilityAdaptive2020}:
\begin{align}
\dot{s}_i(t) &= v_{i-1}(t)-v_i(t), \\
\dot{v}_i(t) &= k_1 [s_i(t-\tau)-\eta-t_h v_i(t)] + k_2 [v_{i-1}(t-\tau)-v_i(t)],
\end{align}
where $s_i(t)$ and $v_i(t)$ are the spacing and velocity of vehicle $i$, $t_h$ is the desired time headway, $\eta$ is the jam spacing, $\tau$ is the control delay, and $k_1$ and $k_2$ are control gains. Parameter values follow Gunter \emph{et al.} and serve as a reproducible empirical proxy for commercially realistic rule-based longitudinal control, rather than a replica of any specific proprietary system. Full parameter values and simulation setup details are provided in Supplementary Section~S4.

\subsection*{Scenario design and energy--emission evaluation}

To quantify the energy and emission consequences of different car-following behaviours under matched traffic conditions, we constructed a 25-vehicle platoon comparison framework consistent with the collective analyses in Figs.~2--4. This platoon size was chosen because it is sufficiently large to capture the downstream propagation and amplification of perturbations, while remaining directly comparable to the spatiotemporal and trajectory analyses used elsewhere in this study. Using the same platoon scale also allows the energy emission results to be interpreted in relation to the observed traffic-wave structure.

For the human-driven baseline, we used measured trajectories from the 25-vehicle Hefei experiment and selected four representative scenarios---\emph{Stable}, \emph{Mild perturbation}, \emph{Strong perturbation}, and \emph{Repeated stop-and-go}---spanning smooth platoon motion through to strongly amplified oscillation (scenario selection criteria and run characteristics are detailed in Supplementary Section~S4). For each scenario, the same leader velocity trajectory, platoon size, and simulation time step were retained for the matched ACC comparison, with following vehicles replaced by automated followers governed by the calibrated ACC model; all followers shared the same parameter set so that differences between HV and ACC arise solely from following behaviour. Metrics were computed over the 24 following vehicles, excluding the leader and an initial warm-up period.

For each vehicle, the speed trajectory $v(t)$ was obtained directly from the measured or simulated trajectories, and the corresponding acceleration trajectory $a(t)$ was estimated using discrete-time differentiation. Fuel consumption was then estimated from the instantaneous tractive power demand (full model parameters and validation in Supplementary Section~S5),
\begin{align}
P_{\mathrm{tr}}(t)
=
mgC_r v(t)
+
\frac{1}{2}\rho C_d A v^3(t)
+
m \max\!\bigl(a(t),0\bigr)\,v(t),
\end{align}
where $m$ is vehicle mass, $g$ is gravitational acceleration, $C_r$ is the rolling-resistance coefficient, $\rho$ is air density, and $C_dA$ is the product of drag coefficient and frontal area. Negative tractive power was truncated to zero, so regenerative braking was not considered.

To account for idling and auxiliary loads, the instantaneous fuel consumption rate was computed as
\begin{align}
\dot{F}(t)
=
\max\!\left(
\dot{F}_{\mathrm{idle}},
\frac{P_{\mathrm{tr}}(t)+P_{\mathrm{aux}}}{\eta H_{\mathrm{fuel}}}
\right),
\end{align}
where $\dot{F}_{\mathrm{idle}}$ is the idle fuel consumption rate, $P_{\mathrm{aux}}$ is the auxiliary power demand, $\eta$ is the powertrain efficiency, and $H_{\mathrm{fuel}}$ is the lower heating value of the fuel.

Integrating $\dot{F}(t)$ over time yields the total fuel consumption of each following vehicle over the analysis window. Summing across all the following vehicles gives the total platoon fuel consumption for that scenario. CO$_2$ emissions were then estimated using a fixed fuel-based emission factor,
\begin{align}
\mathrm{CO}_2 = F \times EF_{\mathrm{CO_2}},
\end{align}
where $F$ is total fuel consumption and $EF_{\mathrm{CO_2}}$ is the CO$_2$ emission factor per unit fuel consumed.

Fuel consumption and CO$_2$ emissions were normalised by vehicle-kilometres travelled. Traffic indicators (mean speed, speed standard deviation, Acc RMS, near-stop ratio) and metric definitions are given in Supplementary Section~S5. Taking the HV platoon as the reference, the relative changes for the ACC platoon were calculated as
\begin{align}
\Delta \mathrm{Fuel}
&=
\frac{\mathrm{Fuel}_{\mathrm{ACC}}-\mathrm{Fuel}_{\mathrm{HV}}}
{\mathrm{Fuel}_{\mathrm{HV}}}
\times 100\%, \\
\Delta \mathrm{CO}_2
&=
\frac{\mathrm{CO}_{2,\mathrm{ACC}}-\mathrm{CO}_{2,\mathrm{HV}}}
{\mathrm{CO}_{2,\mathrm{HV}}}
\times 100\%.
\end{align}
This procedure allows the additional energy and emission costs induced by commercially implemented rule-based automated following to be quantified directly across different perturbation regimes.

\bibliography{Reference20250603}

\section*{Data availability}

The NGSIM trajectory data are publicly available from the US Federal Highway Administration (FHWA) at \url{https://ops.fhwa.dot.gov/trafficanalysistools/ngsim.htm}. The OpenACC database is publicly available at \url{https://data.jrc.ec.europa.eu/dataset/9702c950-c80f-4d2f-982f-44d06ea0009f}~\cite{Makridis2021}. The Hefei 25-vehicle platoon experiment data and the controlled 11-driver car-following data are available from the  authors upon reasonable request at \url{https://traffic-open-data.com/}.

\section*{Code availability}

The analysis code used to process trajectory data, estimate speed--spacing boundaries, fit ex-Gaussian distributions, simulate the ACC platoon, and compute energy--emission metrics is available from the corresponding authors upon reasonable request.

\section*{Acknowledgements}

This work was supported by the National Natural Science Foundation of China 
(Grant Nos. 72222021, 72431006, 72288101, W2411064, and 72401022). 
S.Z.\ acknowledges financial support from the China Scholarship Council 
(CSC; Grant No. 202506250110) through the Joint PhD Training Programme. 
C.J.\ acknowledges support from the UKRI Metascience Research Grants Programme 
(OPP569).

\section*{Author contributions statement}

Shirui Zhou: Writing –- original draft, Writing –- review \& editing, Visualisation, Methodology, Investigation, Formal analysis.  Ching Jin: Writing –- original draft, Writing –- Review \& editing, Data Analysis, Methodology, Visualisation, Data curation. Shiteng Zheng: Experiment, Investigation.  Junfang Tian: Writing – original draft, Writing –- review \& editing,  Supervision, Methodology, Investigation, Funding acquisition, Conceptualisation. Rui Jiang: Writing –- review \& editing, Methodology, Investigation, Funding acquisition, Conceptualisation. Shoufeng Ma: Formal analysis, Funding acquisition, Supervision. Vittorio Loreto: Methodology, Supervision, Writing –- review \& editing.

\section*{Competing Interests}

The authors declare no competing interests.

\end{document}